\documentclass[preprint,twocolumn,5p,times]{elsarticle}

\usepackage{graphicx,epsfig,amssymb,amsmath,amsthm,amscd,amsfonts,bm,accents,mathtools,cuted,hyperref} 

\usepackage{lineno}

\usepackage{float}

\usepackage{pifont}
\usepackage{stmaryrd}
\usepackage{wasysym}
\usepackage{marvosym}

\usepackage{appendix}

\usepackage[T1]{fontenc}
\usepackage{yfonts}
\usepackage{textcomp}
\usepackage{lingmacros}
\usepackage[usenames]{color}

\usepackage[mathscr]{eucal}

\usepackage{natbib}

\DeclareMathAlphabet{\mathpzc}{OT1}{pzc}{m}{it}

\definecolor{HSafron}{RGB}{232,125,30}

\newcommand{\lr}[1]{\left(#1\right)} 
\newcommand{\RB}{Rayleigh-B\'{e}nard }
\newcommand{\vel}{\mathbf{u}} 
\newcommand{\press}{P} 
\newcommand{\res}[1]{#1^{h}} 
\newcommand{\sub}[1]{#1^{\prime}} 
\newcommand{\coor}{\mathbf{x}} 
\newcommand{\pt}[1]{\frac{\partial #1}{\partial \mathrm{t}}} 
\newcommand{\od}[2]{\frac{\mathrm{d} #1}{\mathrm{d #2}}} 
\newcommand{\sol}{\mathbf{U}} 
\newcommand{\solh}{\mathbf{U}^{h}} 
\newcommand{\solp}{\mathbf{U}^{\prime}} 
\newcommand{\wgt}{\mathbf{W}} 
\newcommand{\inner}[2]{\left(#1, #2\right)} 
\newcommand{\V}{\boldsymbol{\mathcal{V}}} 
\newcommand{\force}{\mathbf{F}} 
\newcommand{\tp}{\mathrm{T}} 
\newcommand{\vwt}{\mathbf{w}} 
\newcommand{\pwt}{q} 
\newcommand{\twt}{s} 
\newcommand{\fv}{\mathbf{f}^{\text{V}}} 
\newcommand{\fT}{\mathbf{f}^{T}} 
\newcommand{\yhat}{\widehat{\mathbf{y}}} 
\newcommand{\stabm}{\boldsymbol{\tau}} 
\newcommand{\mt}{\mathbf{G}} 
\renewcommand{\div}{\nabla\cdot} 

\journal{Mech.\ Res.\ Commun.}

\begin{document}

\begin{frontmatter}

\title{High Rayleigh number variational multiscale large eddy simulations of Rayleigh-B\'{e}nard Convection} 

\author[harvard]{David Sondak\corref{cor1}}
\cortext[cor1]{Corresponding author}
\ead{dsondak@seas.harvard.edu}
\address[harvard]{Institute for Applied Computational Science, Harvard University, Maxwell Dworkin, Suite G107, 33 Oxford
Street, Cambridge, MA 02138}

\author[sandia-cs]{Thomas M. Smith}
\author[sandia-cs]{Roger P. Pawlowski}
\address[sandia-cs]{Sandia National Laboratories, Computational Science Department}
\author[sandia-math]{Sidafa Conde}
\address[sandia-math]{Sandia National Laboratories, Computational Mathematics Department}
\author[sandia-math,unm]{John N. Shadid}
\address[unm]{Department of Mathematics and Statistics, University of New Mexico, MSC01 1115, Albuquerque, NM 87131, USA}

\begin{abstract}
  The variational multiscale (VMS) formulation is used to develop residual-based VMS large eddy simulation (LES) models for
\RB convection.  The resulting model is a mixed model that incorporates the VMS model and an eddy viscosity model.  The
Wall-Adapting Local Eddy-viscosity (WALE) model is used as the eddy viscosity model in this work.  The new LES models were 
implemented in the finite element code Drekar.  Simulations are performed using continuous, piecewise linear finite elements.  
The simulations ranged from $Ra=10^6$ to $Ra=10^{14}$ and were conducted at $Pr=1$ and $Pr=7$.  Two domains were considered:  
a two-dimensional domain of aspect ratio $2$ with a fluid confined between two parallel plates and a three-dimensional cylinder 
of aspect ratio $1/4$.  The Nusselt number from the VMS results is compared against three dimensional direct numerical
simulations and experiments.  In all cases, the VMS results are in good agreement with existing literature.
\end{abstract}


\end{frontmatter}


\section{Introduction}
\label{sec:intro}
\RB convection is the buoyancy-driven flow of a fluid confined between two parallel, horizontal plates where the bottom plate
is at a higher temperature than the top plate.  An initially quiescent fluid will be set into motion for a sufficiently 
large temperature difference between the two plates.  This deceptively simple configuration provides for an exceptionally 
rich variety of fluid motion.  Indeed, \RB convection has been used as a proxy for the phenomenon of thermal convection,
which is responsible for a dizzying array of fluid phenomena from the geophysical through the astrophysical.  \RB convection
is also known for being one of the original flow fields studied in the field of hydrodynamic stability
theory~\cite{drazin2004hydrodynamic, chandrasekhar2013hydrodynamic}.  In his pioneering
work, Rayleigh used linear stability theory to show precisely when an initially quiescent fluid will bifurcate from the
quiescent, conduction state to the first convection state~\cite{rayleigh1916lix}.  The primary control parameter governing
this bifurcation is now known as the Rayleigh number $Ra$, which is a measure of the ratio of buoyancy-driven inertial forces
to viscous forces.  Over the years, \RB convection has been studied well beyond the theory of fluid stability.

One major research thrust has been the focus on quantifying how the heat transport through the fluid layer depends on
$Ra$~\cite{bodenschatz2000recent, ahlers2009heat}. 
The primary diagnostic quantity in \RB convection is the dimensionless heat transport expressed as the ratio of total heat 
transport to conduction heat transport and called the Nusselt number ($Nu$).  Significant theoretical, computational, and 
experimental resources have been devoted to determining the relationship between $Nu$ and $Ra$.  A major question focuses on the 
exponent in the power law relationship $Nu\propto Ra^{\beta}$.  Theoretical arguments have been used to show that $\beta\approx
1/3$~\cite{malkus1954heat, howard1966convection} while other rigorous mathematical arguments have established bounds that
show $\beta \leq 1/2$~\cite{doering1996variational,ding2019exhausting}.  Classical work based on
turbulence mixing length models has predicted that $\beta$ transitions to $1/2$ for very large $Ra$ with logarithmic corrections in
$Ra$~\cite{kraichnan1962turbulent}.  
Other recent work has proposed models of $Nu$ that are not pure power laws in $Ra$~\cite{grossmann2000scaling}.  There has
been much discussion on recent experimental results at high $Ra$ that observe a transition to
$\beta=1/2$~\cite{he2012transition,bouillaut2019transition}
or not at very high $Ra$~\cite{niemela2000turbulent,doering2019thermal}.  Numerical calculations have shown that
$\beta\approx 0.28 - 0.3$ up to the largest $Ra$ currently
achievable~\cite{stevens2011prandtl,iyer2020classical}.  Very recently, two-dimensional numerical simulations up to
$Ra=10^{14}$ observed a transition to $\beta=1/2$~\cite{zhu2018transition,zhu2019absence}, while three-dimensional
simulations up to $Ra=10^{15}$ have not observed this transition~\cite{iyer2020classical}.  At high $Ra$, it becomes
prohibitively expensive to perform fully-resolved direct numerical simulations of \RB convection.  Hence, there is great 
interest in the development of turbulence models that will permit accurate simulations at high $Ra$.

Instead of directly resolving all scales of motion, large eddy simulation (LES) coarse grains the fields and equations and 
simulates only the largest scales of motion.  The price to pay is that this coarse graining procedure introduces correlations 
between resolved and unresolved terms that cannot be neglected and must be modeled.  The goal of LES turbulence modeling is 
to develop models that account for the effect of the unresolved scales on the resolved scales.  Fluid simulations with the 
finite element method have been challenging due to the need to satisfy (or circumvent) the inf-sup condition, satisfy the
incompressibility constraint, and stabilize spurious oscillations for highly convective flows.  Stabilized finite element 
methods were developed to overcome these challenges~\cite{brooks1982streamline,franca2004stabilized} and were eventually
shown to derive from the variational multiscale (VMS) method~\cite{hughes1995multiscale}.  Since its original development,
the VMS method has been used to develop LES models for a variety of fluid
flows~\cite{masud2006multiscale,bazilevs2007variational,liu2012residual,sondak2015new,codina2000stabilized,codina2018variational}.  
Recently, researchers developed a VMS-based LES model for \RB convection with application to heating
systems~\cite{xu2019residual}.  In the current work, we propose a mixed VMS method for \RB convection at high $Ra$.  We perform 
simulations up to $Ra=10^{14}$ for two different Prandtl numbers ($Pr=1$ and $Pr=7$) in both two and three dimensions for
rectangular and cylindrical geometries.

The remainder of the paper is organized as follows.  In section~\ref{sec:bg} we provide the governing equations, the VMS
formulation for \RB convection, and a description of the code used to perform the simulations.  Following this,
section~\ref{sec:results} presents the results of the simulations.  Section~\ref{sec:conclusions} summarizes the work and
discusses ongoing and future work.

\section{Background}
\label{sec:bg}

\subsection{Rayleigh-B\'{e}nard Convection}
\label{sec:gov-eqns}
\RB convection is concerned with the buoyancy-driven flow of a fluid confined between two parallel, horizontal plates
separated by a distance $H$.  The two plates are maintained at constant temperatures such that the temperature difference
between the top and bottom plates is $\Delta T = T_{\text{bot}} - T_{\text{top}} > 0$.  Within the Oberbeck-Boussinesq
approximation, density variations are assumed to be important only in the buoyancy term and these variations are taken to
depend linearly on the temperature.  The fluid is otherwise incompressible.  The velocity field $\vel\lr{\coor,t}=\lr{u,v,w}$ 
evolves according to the Oberbeck-Boussinesq equations,
\begin{align}
  \rho_{0}\lr{\pt{\vel} + \div\lr{\vel\otimes\vel}} &= -\nabla \press + \mu\nabla^{2}\vel +\alpha_{V}g\lr{T - T_{0}}\widehat{\mathbf{y}} \label{eq:mom} \\
  &\div\vel = 0 \label{eq:inc}
\end{align}
where $\rho_{0} = \rho\lr{T_{0}}$ is the reference density evaluated at a reference temperature $T_{0}$,
$\press=\press\lr{\coor,t}$ is the kinematic pressure of the fluid, $\mu$ is the kinematic viscosity, $\alpha_{V}$ is the
coefficient of volume 
expansion of the fluid, $g$ is the acceleration due to gravity, and $\yhat$ is the unit vector in the vertical 
direction.  The temperature field $T=T\lr{\coor,t}$ evolves according to an advection diffusion equation,
\begin{align}
  \rho_{0}C_{p}\lr{\pt{T} + \div\lr{\vel T}} = k\nabla^{2}T \label{eq:temp}
\end{align}
where $C_{p}$ and $k$ are the specific heat and thermal conductivity of the fluid, respectively.  The velocity field uses
no-slip boundary conditions on all solid surfaces.  The temperature is held at a uniform constant temperature on the top and
bottom surfaces such that the bottom surface is hotter than the top surface.  In the present work, we consider two
different geometries and therefore the boundary conditions on the ``sides'' are specified differently depending on which
geometry is being considered.  In two-dimensional \RB convection between two infinite parallel
planes the velocity and temperature fields have periodic boundary conditions in the $x$ direction.  In three-dimensional \RB convection in a
right circular cylinder, the surface of the cylinder is insulated and the temperature field uses homogeneous Neumann boundary 
conditions on the side-walls of the cylinder.

The two classical non-dimensional parameters emerging from the system in~\eqref{eq:mom}-~\eqref{eq:temp} are the Rayleigh and
Prandtl numbers.  The Prandtl number is a fluid property and is given by,
\begin{align}
  Pr = \frac{\nu}{\kappa}
\end{align}
where $\nu = \mu/\rho_{0}$ and $\kappa = k / \lr{\rho_{0} C_{p}}$.  The Rayleigh number is, 
\begin{align}
  Ra = \frac{g\alpha_{V}\Delta T H^{3}}{\nu\kappa}
\end{align}
and is interpreted as a measure of the strength of buoyancy-driven inertial forces.  In the
conduction state, the heat transport is $\mathcal{H}_{\text{cond}} = \kappa\Delta T / H$, independent of $Ra$ and $Pr$.
After convection sets in, the heat transport is quantified by the Nusselt number $Nu$ as the ratio of total heat transfer to 
conduction heat transfer.  The Nusselt number is, 
\begin{align}
  Nu = -\frac{H}{\Delta T}\overline{\left.\od{T}{y}\right|_{y_{\text{wall}}}} \label{eq:Nu1}
\end{align}
where $\overline{\ \cdot \ }$ represents an average over the plane orthogonal to the wall-normal coordinate.  In
statistically steady state, after integrating across the width of the fluid layer, the Nusselt number can be written as, 
\begin{align}
  Nu = 1 + \frac{\left<vT\right>}{\mathcal{H}_{\text{cond}}} \label{eq:Nu2}
\end{align}
where $\left<\cdot\right>$ is a space-time average.  Once a statistically stationary state has been reached, the time-average 
of~\eqref{eq:Nu1} is equal to~\eqref{eq:Nu2}.  In the present work, the Nusselt number was calculated using~\eqref{eq:Nu1}
and~\eqref{eq:Nu2} with identical results.

\subsection{Variational Multiscale Formulation for Rayleigh-B\'{e}nard Convection}
\label{sec:vms}
The variational statement of the equations governing Rayleigh-B\'{e}nard convection is:  \textit{Find } $\sol\in\V$
\textit{s.t.} $\ \forall \ \wgt\in\V$ 
\begin{align}
  \mathcal{A}\inner{\wgt}{\sol} = \inner{\wgt}{\force} \label{eq:varstat}
\end{align}
where $\sol=\left[\vel, \press, T\right]^{\tp}$ is a vector of solutions, $\wgt=\left[\vwt, \pwt, \twt\right]^{\tp}$ is a
vector of weighting functions, and $\force=\left[\fv, 0, \fT\right]^{\tp}$ is a vector of forcing functions.  Note that in
the current work this forcing is zero.  As per usual convection, the notation $\inner{\cdot}{\cdot}$ denotes an $L_{2}$ inner
product of two functions over the domain $\Omega$.  The semilinear form~\eqref{eq:varstat} is,
\begin{align}
  \mathcal{A}\inner{\wgt}{\sol} = \mathcal{A}^{V}\inner{\wgt}{\sol} + \inner{q}{\div\vel} +
\mathcal{A}^{T}\inner{\wgt}{\sol}
\end{align}
where 
\begin{align}
 \mathcal{A}^{V}\inner{\wgt}{\sol} &= \inner{\vwt}{\rho_{0}\pt{\vel}} - \inner{\nabla\vwt}{\rho_{0}\vel\otimes\vel} -
\inner{\div\vwt}{P} \nonumber \\ 
 &\hspace{1.0em}+ \inner{\nabla\vwt}{\mu\nabla\vel} - \inner{\vwt}{\alpha_{V}g\left(T - T_{0}\right)\yhat} \\
 \mathcal{A}^{T}\inner{\wgt}{\sol} &= \inner{\twt}{\rho_{0}C_{p}\pt{T}} - \inner{\nabla\twt}{\rho_{0}C_{p}\vel T} \nonumber
\\ &\hspace{1.0em}+ \inner{\nabla\twt}{k\nabla T}.
\end{align}
No boundary terms appear in the variational formulation due to the periodic, Dirichlet, and homogeneous Neumann boundary
conditions in the problems considered in this work.

We consider a finite element method in which the discretized solutions $\sol^{h}\in\V^{h}\subset\V$ are linear combinations of
bilinear quadrilateral or hexahedral basis functions.  The straightforward discretization leading to the Galerkin statement:
\textit{Find} $\sol^{h}\in\V^{h}$ \textit{s.t.} $\ \forall \ \wgt^{h}\in\V^{h}, \ \ \mathcal{A}\inner{\wgt^{h}}{\sol^{h}} =
\inner{\wgt^{h}}{\force}$ 
is not sufficient due to the instabilities inherent in the Galerkin method for highly convective flows.  To
overcome this limitation, we develop a mixed variational multiscale formulation for Rayleigh-B\'{e}nard convection.  The VMS method
induces a sum-decomposition of the solution field $\sol$ into resolved $\solh$ and unresolved $\solp$ components so that
$\sol = \solh + \solp$.  The resulting VMS formulation for our problem with linear finite elements is:  \textit{Find
} $\sol^{h}\in\V^{h}$ \textit{s.t.} $\ \forall \ \wgt^{h}\in\V^{h}$
\begin{align}
  &\mathcal{A}\inner{\wgt^{h}}{\sol^{h}} - \underbrace{\inner{\nabla\res{\vwt}}{\rho_{0}\res{\vel}\otimes\sub{\vel}}}_{\text{VMS
cross stresses}} -
\underbrace{\inner{\nabla\res{\vwt}}{\rho_{0}\sub{\vel}\otimes\res{\vel}}}_{\text{SUPG}} \nonumber \\
&\hspace{3.0em}-\underbrace{\inner{\nabla\res{\vwt}}{\rho_{0}\sub{\vel}\otimes\sub{\vel}}}_{\text{Reynolds stresses}}
\nonumber - \inner{\div\res{\vwt}}{\sub{\press}} - \inner{\res{\vwt}}{\alpha_{V}g\sub{T}\yhat} \nonumber \\
&\hspace{1.5em}-\inner{\nabla\res{q}}{\sub{\vel}} \label{eq:RBC_VMS} \\ 
  &\hspace{1.5em}- \underbrace{\inner{\nabla\res{\twt}}{\rho_{0}C_{p}\res{\vel}\sub{T}}}_{T \text{ SUPG}} -
\underbrace{\inner{\nabla\res{\twt}}{\rho_{0}C_{p}\sub{\vel}\res{T}}}_{T \text{ VMS cross stresses}} \nonumber \\ 
&\hspace{3.0em}- \underbrace{\inner{\nabla\res{\twt}}{\rho_{0}C_{p}\sub{\vel}\sub{T}}}_{T \text{ Reynolds stresses}} \nonumber \\
  &\hspace{1.5em}= \inner{\res{\wgt}}{\force}. \nonumber  
\end{align}
The formulation in~\eqref{eq:RBC_VMS} neglects terms involving time derivatives of unresolved fields as well as inner
products of gradients of resolved and unresolved fields.  Although not pursued here, approaches exist to model the transient
effects of the unresolved scales~\cite{codina2002stabilized}.  In residual-based VMS formulations, the unresolved fields are
proportional to the residual of the partial differential equations (PDEs),
\begin{align}
  \solp \approx -\stabm\boldsymbol{\mathcal{R}}\lr{\solh} \label{eq:uprime}
\end{align}
where 
\begin{strip}
\begin{align}
  \boldsymbol{\mathcal{R}}\lr{\solh} = 
    \begin{bmatrix}
      \displaystyle \rho_{0}\pt{\res{\vel}} + \rho_{0}\div\lr{\res{\vel}\otimes\res{\vel}} + \nabla \res{\press}
-\mu\nabla^{2}\res{\vel} - \alpha_{V}g\lr{\res{T} - T_{0}}\yhat  \\
      \div\res{\vel}  \\
      \displaystyle \rho_{0}C_{p}\lr{\pt{\res{T}} + \div\lr{\res{\vel} \res{T}}} - k\nabla^{2}\res{T}
    \end{bmatrix}
\end{align}
\end{strip}
and $\stabm$ is the stabilization matrix.  We use a diagonal stabilization matrix $\stabm = \text{diag}\lr{\tau^{V}_{ii}, \tau^{C}, \tau^{T}}$ 
where
\begin{strip}
\begin{align}
  \tau^{V}_{ii} &= \left[\lr{\dfrac{2C_{t}^{V}\rho_{0}}{\Delta t}}^{2} + \rho_{0}^{2}\res{\vel}\cdot\mt\res{\vel} +
\lr{C_{1}\mu}^{2}\|\mt\|^{2} + \rho_{0}^{2}\alpha_{V}g\|\res{T}\|\|\mt\|^{1/2}\right]^{-1/2}, \quad i=1,\ldots,n_{sd} \\
  \tau^{C} &= \lr{C_{t}^{V}\text{trace}\lr{\mt}\tau^{V}}^{-1} \\
  \tau^{T} &= \left[\lr{\dfrac{2C_{t}\rho_{0} C_{p}}{\Delta t}}^{2} + \lr{\rho_{0} C_{p}}^{2}\res{\vel}\cdot\mt\res{\vel} +
\lr{C_{1}k}^{2}\|\mt\|^{2}\right]^{-1/2},
\end{align}
\end{strip}
$\mt$ is the metric tensor, $n_{sd}$ the number of spatial dimensions, $\Delta t$ is the time-step, and $C_{1} = C_{t} = 1$.
The components of the metric tensor $\mt$ are given by,
\begin{align}
  G_{ij} = \frac{\partial\xi_{k}}{\partial x_{i}}\frac{\partial\xi_{k}}{\partial x_{j}}
\end{align}
where $\boldsymbol{\xi}$ are the coordinates in the parametric (finite element) space.
The first-order approximation to the unresolved scales~\eqref{eq:uprime} has been shown to be insufficient to model
correlations of unresolved scales (the Reynolds stresses, $\sub{\vel}\otimes\sub{\vel}$ and $\sub{\vel}\sub{T}$) for highly
turbulent flows~\cite{wang2010spectral}.  We expect the Reynolds stresses to play a role in high $Ra$ \RB convection.  One
option to more accurately model the Reynolds stress terms is to work with higher-order methods~\cite{hughes2005isogeometric,
bazilevs2007variational}.  An alternative approach is to introduce a mixed model wherein the Reynolds stresses are modeled by
a classical eddy viscosity model (EVM)~\cite{wang2010mixed, oberai2014residual, sondak2015new}.  The mixed-model for \RB
convection is:  \textit{Find} $\sol^{h}\in\V^{h}$ \textit{s.t.} $\ \forall \ \wgt^{h}\in\V^{h}$
\begin{align}
  &\mathcal{A}\inner{\wgt^{h}}{\sol^{h}} - C_{\text{VMS}}^{V}\inner{\nabla\res{\vwt}}{\rho_{0}\res{\vel}\otimes\sub{\vel}} -
C_{\text{SUPG}}^{V}\inner{\nabla\res{\vwt}}{\rho_{0}\sub{\vel}\otimes\res{\vel}} \nonumber \\ 
  &\hspace{1.5em}- C_{P}^{V}\inner{\div\res{\vwt}}{\sub{\press}} + C_{\text{B}}^{V}\inner{\res{\vwt}}{\alpha_{V}g\sub{T}\yhat} -
C_{\text{PSPG}}^{V}\inner{\nabla\res{q}}{\sub{\vel}} \nonumber \\ 
  &\hspace{1.5em}- C_{\text{VMS}}^{T}\inner{\nabla\res{\twt}}{\res{\vel}\sub{T}} - C_{\text{SUPG}}^{T}\inner{\nabla\res{\twt}}{\sub{\vel}\res{T}} -
\label{eq:RBC_MIXED} \\
  &\hspace{1.5em}+ C_{\text{EVM}}^{V}\inner{\nabla\res{\vwt}}{\nu_{T}\nabla^{s}\res{\vel}} +
C_{\text{EVM}}^{T}\inner{\nabla\res{\twt}}{\rho_{0}C_{p}\kappa_{T}\nabla\res{T}} \nonumber \\
  &\hspace{1.5em}= \inner{\res{\wgt}}{\force} \nonumber
\end{align}
where $\nabla^{s}\vel = \lr{\nabla\vel + \lr{\nabla\vel}^{\tp}}/2$.
In~\eqref{eq:RBC_MIXED}, the first term on the first line represents the Galerkin discretization, the second term represents
a VMS cross-stress term, and the third term represents the SUPG stabilization.  The first term on the second line is the
pressure stabilization term, the second term represents temperature fluctuations in the buoyancy term, and the third term is
used to overcome the inf-sup condition for finite element discretizations of the incompressible Navier-Stokes equations.  The
first term on the third line is an upwinding stabilization term for the temperature advection-diffusion equation and the
second term is an additional cross-stress term from the VMS formulation.  Finally, the fourth line includes the EVMs that are
used to model the Reynolds stress terms.  We have also included coefficients before each term which can be used to toggle
various models on and off.  The simulations in the current study are performed with two versions of~\eqref{eq:RBC_MIXED}.
The first version is a straight VMS implementation neglecting the Reynolds stresses in addition to the other neglected terms
mentioned above.  The second version uses a modified VMS formulation and incorporates the Wall-Adapting Local Eddy-viscosity
(WALE) model~\cite{nicoud1999subgrid}. The WALE model possesses several desirable properties including that $\nu_{T}$
naturally approaches zero near the wall.  In classical wall-bounded flows, the WALE model also recovers the correct
asymptotic behavior of $\nu_{T}$ near the wall, $\nu_{T}\sim y^{3}$ for $y\to 0$.  The VMS-WALE mixed model in the current work
uses the SUPG stabilization terms as well as the VMS terms for the velocity-pressure saddle point system.  In this way, the
VMS-WALE model corresponds to a classical finite element WALE simulation with the necessary stabilization terms.
Table~\ref{tab:mixed_params} provides a summary of the toggling coefficients used in this study.
\begin{table}
  \begin{center}
\def~{\hphantom{0}}
  \begin{tabular}{ccc}
       Coefficient           & VMS Model & VMS-WALE \\[3pt]
       $C_{\text{VMS}}^{V}$  & 1         & 0        \\[2pt]
       $C_{\text{SUPG}}^{V}$ & 1         & 1        \\[2pt]
       $C_{P}^{V}$           & 1         & 1        \\[2pt] 
       $C_{\text{B}}^{V}$    & 1         & 0        \\[2pt]
       $C_{\text{PSPG}}^{V}$ & 1         & 1        \\[2pt]
       $C_{\text{VMS}}^{T}$  & 1         & 0        \\[2pt]
       $C_{\text{EVM}}^{V}$  & 0         & 1        \\[2pt]
       $C_{\text{EVM}}^{T}$  & 0         & 1        \\[2pt]
  \end{tabular}
  \caption{Selected parameters in the mixed VMS-EVM model \eqref{eq:RBC_MIXED} for the two models used in the current study.}
  \label{tab:mixed_params}
  \end{center}
\end{table}
As a final remark, we note that when using linear finite elements, the diffusive terms in the PDE residual are identically
zero.  In the \RB convection problem, these diffusive fluxes may play a key role and neglecting them in the residual may have
a negative impact on the final solution~\cite{jansen1999better}.  We leave the development and implementation of such models
for future work.

\subsection{Numerical Methodology}
\label{sec:numerics}
All simulations were conducted using the Drekar finite element code~\cite{shadid-jcp-10_mhd,ShadidetalResistiveMHD2016} using
linear quadrilateral or hexahedral finite elements.  The 2D simulations used an SDIRK22 (singly diagonally implicit
Runge-Kutta 2nd order, 2 stage) time-integration method while the 3D simulations used a second order BDF method.  The
two-dimensional computations were run on up to 2000 cores while the three-dimensional simulations were run on up to 8000
cores.  All simulations were performed on a capacity cluster with an Intel Haswell based CPU.  The following section provides
a broad overview of the time integration and solver formulations used to perform the simulations.

\subsubsection{Fully-implicit Time Integration and Strongly-coupled Newton-Krylov-AMG Solver} 
\label{subsubsec:ImplicitNK}
Fully-implicit formulations, coupled with effective robust nonlinear iterative solution methods,
have the potential to provide stable, higher-order time-integration of multiphysics systems
when long dynamical time-scales are of interest. These methods can
follow the desired dynamical time-scales
as opposed to time-scales determined by either numerical stability or by temporal order-of-accuracy reduction~\cite{OBER02,ROPP03,Ropp05,KnollWave05,Ropp08,shadid-jcp-10_mhd,ShadidetalResistiveMHD2016}. 
For time-integration of the governing equations in \eqref{eq:mom} - \eqref{eq:temp},
the  L-stable SDIRK22 method is used~\cite{shadid-jcp-10_mhd,ShadidetalResistiveMHD2016} as it provides 
a high-order time integration with damping of the highest unresolved wavenumbers~\cite{AscherPetzold98}.  

A finite element discretization of the VMS equations~\eqref{eq:RBC_VMS}
gives rise to a system of coupled, nonlinear, nonsymmetric algebraic
equations, the numerical solution of which can be very
challenging. These equations are linearized using an inexact form of Newton's method.
A formal block matrix representation of these discrete linearized equations is given by
\begin{align*}
\begin{bmatrix}
{\bf D_u} & {\bf B^\tp} & {\bf Q}  \\
{\bf B} & {\bf L_P} & {\bf 0}  \\
{\bf C} & {\bf 0} & {\bf D}_T
\end{bmatrix} 
\left[
\begin{array} {c}
{\bf  \delta \hat u} \\
{\bf \delta \hat P} \\
{\bf \delta \hat T} \\
\end{array}
\right]
=
-\left[
\begin{array} {c}
{\bf F_u} \\
{\bf F_P} \\
{\bf F_T} 
\end{array}
\right].
\label{eqn:4x4BlockMatrix}
\end{align*}
where the block diagonal contribution of the
stabilization procedure has been highlighted by a specific ordering.
The block matrix, ${\bf D_u}$, corresponds to the discrete transient, convection, diffusion and stress terms
acting on the unknowns ${\bf \delta \hat u}$; the matrix
${\bf B}^\tp$ corresponds to the discrete gradient operator; ${\bf B}$,
the divergence operator; the block matrix ${\bf D}_T$
corresponds to the discrete transient, convection, diffusion operator acting on the temperature, and the matrix
${\bf L_P}$  corresponds to the discrete ``pressure
Laplacian''. The matrix ${\bf Q}$ corresponds to the coupling of velocity and the temperature field (buoyancy term), and the matrix ${\bf C}$ the coupling of temperature gradient to the velocity field.  
The right hand side vectors  contain the residuals for Newton's method.
The existence of the nonzero matrix
${\bf L_P}$ in the stabilized finite element discretization, is in contrast to Galerkin methods for incompressible flow using mixed interpolation that
produce a zero block on the total mass continuity diagonal.
The existence of this block matrix  helps to enable the solution
of the linear systems with a number of algebraic, domain
decomposition (DD), and algebraic multilevel (AMG) type preconditioners/smoothers
 that rely on non-pivoting ILU type factorization,
or in some cases methods such as Jacobi or Gauss-Seidel as
sub-domain solvers \cite{shadid-jcp-10_mhd,ShadidetalResistiveMHD2016}. 
Although the above formal block matrix representation provides insight into the system,
the actual linear algebra implementation in the application employs an ordering
by finite element mesh node with each degree of freedom (dof) ordered consecutively. The Jacobian is evalauted analytically using automatic differentiation \cite{phipps2012ad}.

Fully-coupled Newton-Krylov techniques~\cite{BS90} where a Krylov solver is used
to solve the linear system generated by a Newton's method are robust.  However, efficient solution of the
large sparse linear system that must be solved for each nonlinear iteration is challenging~\cite{Knoll-Keyes,Lin06}.
The performance and scalability of the preconditioner is critical~\cite{Knoll-Keyes}.
It is well known in the literature that Schwarz DD preconditioners do not scale due to lack of global coupling~\cite{smith-ddbook}.
Multigrid methods are one of the most efficient techniques for solving large linear systems~\cite{Oosterlee}.
A Newton-Krylov preconditioned by AMG solution method has been described in our previous work in
detail~\cite{Lin06,lin2009jcp,shadid-jcp-10_mhd,ShadidetalResistiveMHD2016,LinShadidetalJCAM2017} and we will therefore only
provide a very brief description here.

A  Newton-Krylov (NK) method is an implementation of Newton's 
method in which a Krylov iterative solution technique is used to approximately 
solve the linear systems, $\bf{J}_k\bf{s}_{k+1}=-\bf{F}_k$, that are generated at 
each step of Newton's method. 
For efficiency, an inexact Newton method~\cite{Dennis-Schnabel,EW96,JN99} is 
usually employed, whereby one approximately solves the linear systems generated 
in the Newton method by choosing a forcing term $\eta_k$ and stopping the Krylov 
iteration when the inexact Newton condition, 
$ \|\bf{F}_k+\bf{J}_k\bf{s}_{k+1}\|\leq\eta_{k+1}\|\bf{F}_k\| $ is satisfied. The particular 
Krylov method that is used in this study is a robust non-restarted GMRES method
that is capable of iteratively converging to the solution of very large non-symmetric 
linear systems 
provided a sufficiently robust and scalable preconditioning method is available~\cite{Lin06,lin2009jcp,shadid-jcp-10_mhd,ShadidetalResistiveMHD2016,LinShadidetalJCAM2017}.
Two nonlinear convergence criteria are used to ensure that
the numerical solution error is below discretization error.  The first
is a sufficient reduction in the relative nonlinear residual norm, $\| F_{k} \| / \|
F_{o} \| < 10^{-2}$.   In general, and specifically in the results presented in this paper, this requirement is easily satisfied.
The second convergence criterion is based on a sufficient decrease of
a weighted norm of the Newton update vector.  This latter
criterion requires that the correction, $\Delta \chi _{i}^{k}$, for any
variable, $\chi _{i}$, is small compared to its magnitude, $\left|\chi
_{i}^{k}\right|$, and is given by
\[
\sqrt{\frac{1}{N_{u}}\sum _{i=1}^{N_{u}}\left[\frac{\left|\Delta \chi _{i}\right|}{\varepsilon _{r}\left|\chi _{i}\right|+\varepsilon _{a}}\right]^{2}}<1,
\]
where $N_{u}$ is the total number of unknowns, $\varepsilon _{r}$ is
the relative error tolerance between the variable correction and its
magnitude, and $\varepsilon _{a}$ is the absolute error tolerance of
the variable correction.  Essentially $\varepsilon _{a}$
sets the magnitude of components that are to be considered
to be numerically zero. 
In the numerical results that are presented in this paper the relative-error, and
absolute-error tolerance are set to $10^{-3}$ and $10^{-6}$ respectively for all of the test cases.  
In general, each linear system in Newton's method is solved to a moderate level of accuracy (e.g. $\eta = 10^{-3}$) since the outer nonlinear 
Newton iteration controls convergence at each time step. 

A scalable preconditioner for the iterative linear solver is necessary to achieve
solutions efficiently.  For this reason,
a fully-coupled algebraic multigrid method is employed~\cite{ml-guide}. 
In general AMG methods
are significantly easier to implement and integrate with complex unstructured mesh 
discretizations than
geometric multigrid methods~\cite{tuminaro2002,shadid2004,shadid2006}.
Our fully-coupled AMG preconditioner employs a nonsmoothed aggregation approach with uncoupled aggregation.
For systems of partial differential equations (PDEs), aggregation is performed on the graph where all the PDEs per mesh node
are represented by a single vertex, as in our VMS discretization of the governing equations with each dof ordered at each
finite element node consecutively.
 The discrete equations are projected to the coarser level employing a Galerkin projection with a triple matrix product,
$A_{\ell+1} = R_\ell A_\ell P_\ell$, where $R_\ell$ restricts the residual from level $\ell$ to level
$\ell\hskip -.035in+\hskip -.035in1$,
$A_{\ell}$ is the discretization matrix on level $\ell$
and $P_\ell$ prolongates the correction from level
$\ell\hskip -.035in+\hskip -.035in1$ 
to $\ell$.
We typically employ both pre- and post-smoothing on each level of the multigrid V-cycle except the coarsest level where a serial sparse direct solve is performed. In the computations carried out in this study the AMG preconditioner employs a DD-ILU(k) smoother with one level of overlap and
a moderate level of fill-in (e.g. k = 1) 
A  detailed discussion of the scalability of the solvers and  comparisions of differing preconditioning techniques is out of scope of this study.
However these fully-coupled Newton-Krylov-AMG solver have been studied extensively, with  results demonstrating scaling on 
up to 1M+ cores for challenging resistive MHD type problems~\cite{ShadidetalResistiveMHD2016,LinShadidetalJCAM2017}.

\section{Results}
\label{sec:results}
VMS-based large eddy simulations of \RB convection were run in two- and three-dimensional domains.  In both cases, the no-slip
boundary condition was used for the velocity field on all surfaces while the temperature field was prescribed at the top and 
bottom surfaces.  Statistics such as the Nusselt number were obtained after the simulations reached a statistically steady
state by averaging over a number of free-fall times, $t_{f} = H / U_{f}$ where the free-fall velocity is given by,
\begin{align}
  U_{f} = \sqrt{g\alpha_{V}\Delta T H}.
\end{align}

\subsection{VMS Simulations of Two-dimensional \RB Convection}
\begin{figure}[h!]
  \centering
  \includegraphics[width=0.45\textwidth]{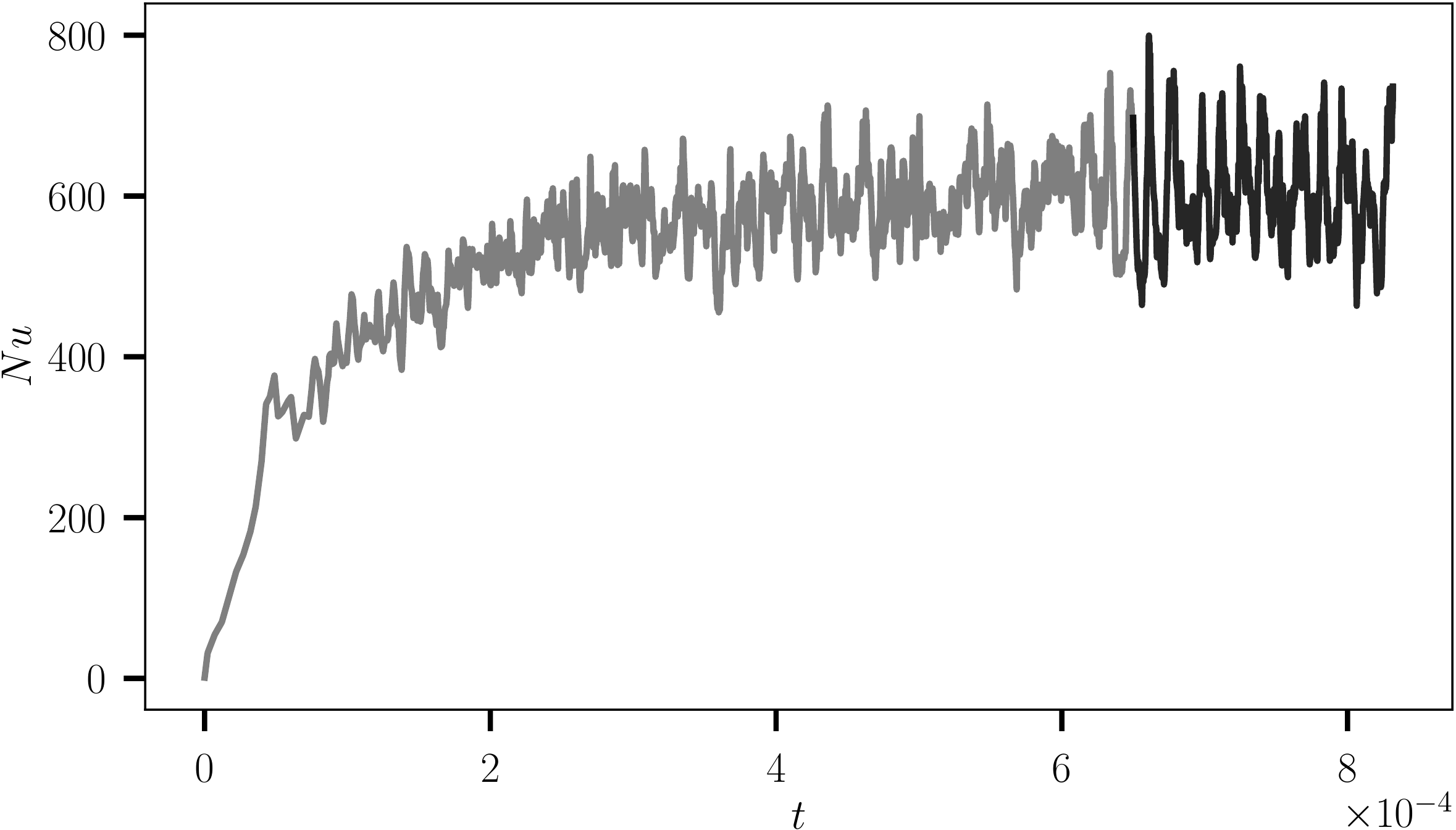}
  \caption{Nusselt evolution showing development to the statistically steady state.  Statistics were collected in the
statistically stationary region, indicted here by the darkly-shaded region.}
  \label{fig:Nu_t}
\end{figure}

\begin{figure}[h!]
  \centering
  \includegraphics[width=0.45\textwidth]{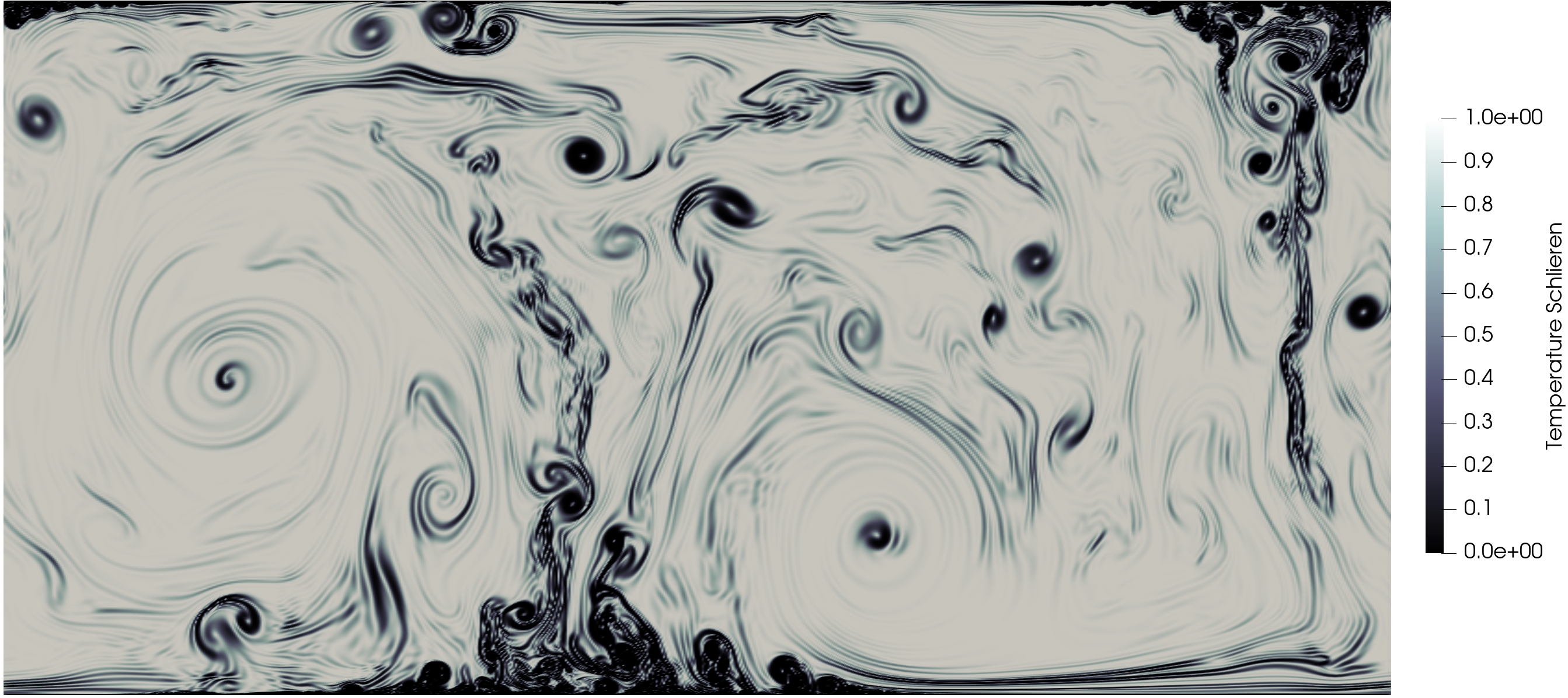}
  \caption{Temperature snapshot of the $Ra=10^{13}$ simulation using a Schlieren-type visualization.}
  \label{fig:Ra10T_snapshot}
\end{figure}

Two-dimensional simulations were run at $Pr=1$.  The aspect ratio for the two-dimensional problem was equal to $2$ and
periodic boundary conditions were used in the streamwise ($x$) direction.  Two types of meshes were used for the 2D 
simulations.  The first mesh used a uniform discretization in both coordinate directions.  The second type of mesh was
uniform in the $x-$ direction but stretched in the $y-$ direction.  The stretching was accomplished using, 
\begin{align}
  y_{j}^{s} = \frac{H}{2}\lr{1 - \cos\lr{\frac{\lr{y_{j}-y_{B}}\pi}{H}}} + y_{B}, \quad j=0,\ldots,N_{y}
\end{align}
where $y_{j}=j\Delta y + y_{B}$, $\Delta y$ is a uniform mesh spacing, $N_{y}$ is the number of elements in the wall-normal
direction, and $y_{B}$ and $y_{T}$ are the bottom and top coordinates of the plates, respectively.  For all of the
two-dimensional simulations, $y_{T}=1$ and $y_{B}=-1$ leading to $H=2$.  We note that, especially for the high $Ra$ cases,
the simulations may be under-resolved in the boundary layer.  This may have significant implications for the $Nu$ results.
Even so, the residual-based LES models do provide some measure of robustness and adaptivity in the near-wall region.  For
$Ra=10^{13}$, we found $y^{+}<1$, which is an indication that the mesh is resolved.  However, a more rigorous mesh convergence
study should be performed.  All statistics were measured after the simulations reached a statistically steady state.  A
representative temporal evolution of $Nu$ is presented in Figure~\ref{fig:Nu_t} for $Ra=10^{13}$.  The lightly-shaded region
represents the transient portion and transition to a statistically steady state while the darkly-shaded region depicts the
period within which statistics were collected.
A representative snapshot of the temperature field from the statistically steady portion of the simulation is presented in 
Figure~\ref{fig:Ra10T_snapshot}.  The visualization uses a Schlieren-type coloring to bring out the features of the flow.
Table~\ref{tab:2D_sim_params} presents a summary of the simulation parameters and results for the 2D VMS runs.
\begin{table}
  \begin{center}
\def~{\hphantom{0}}
  \begin{tabular}{cccccc}
      $Ra$           & $Nu$      & $N_{x}$ & $N_{y}$ & $n_{f}$ & Mesh        \\[3pt]
       $10^6$        & 8.21      & 128     & 128     & 2250    & Uniform     \\
       $10^7$        & 13.4      & 256     & 256     & 3160    & Non-uniform \\
       $10^8$        & 24.9      & 512     & 256     & 170     & Uniform     \\
       $2\cdot 10^8$ & 28.3      & 512     & 256     & 664     & Uniform     \\
       $10^9$        & 43.1      & 512     & 256     & 761     & Non-uniform \\
       $10^{10}$     & 89.8      & 512     & 256     & 1480    & Non-uniform \\
       $10^{11}$     & 175       & 512     & 256     & 644     & Non-uniform \\
       $10^{12}$     & 347       & 650     & 350     & 192     & Non-uniform \\
       $10^{13}$     & 601       & 750     & 500     & 144     & Non-uniform \\
       $10^{14}$     & 1172      & 750     & 500     & 139     & Non-uniform \\
  \end{tabular}
  \caption{Simulation parameters and results for the two-dimensional runs.  $N_{x}$ and $N_{y}$ are the number of linear
finite elements used for spatial discretization.  $n_{f}$ represents the number of free-fall times over which statistics were
computed.  The last column indicates if a uniform or stretched mesh was used in the wall-normal direction.}
  \label{tab:2D_sim_params}
  \end{center}
\end{table}

\begin{figure}[h!]
  \centering
  \includegraphics[width=0.45\textwidth]{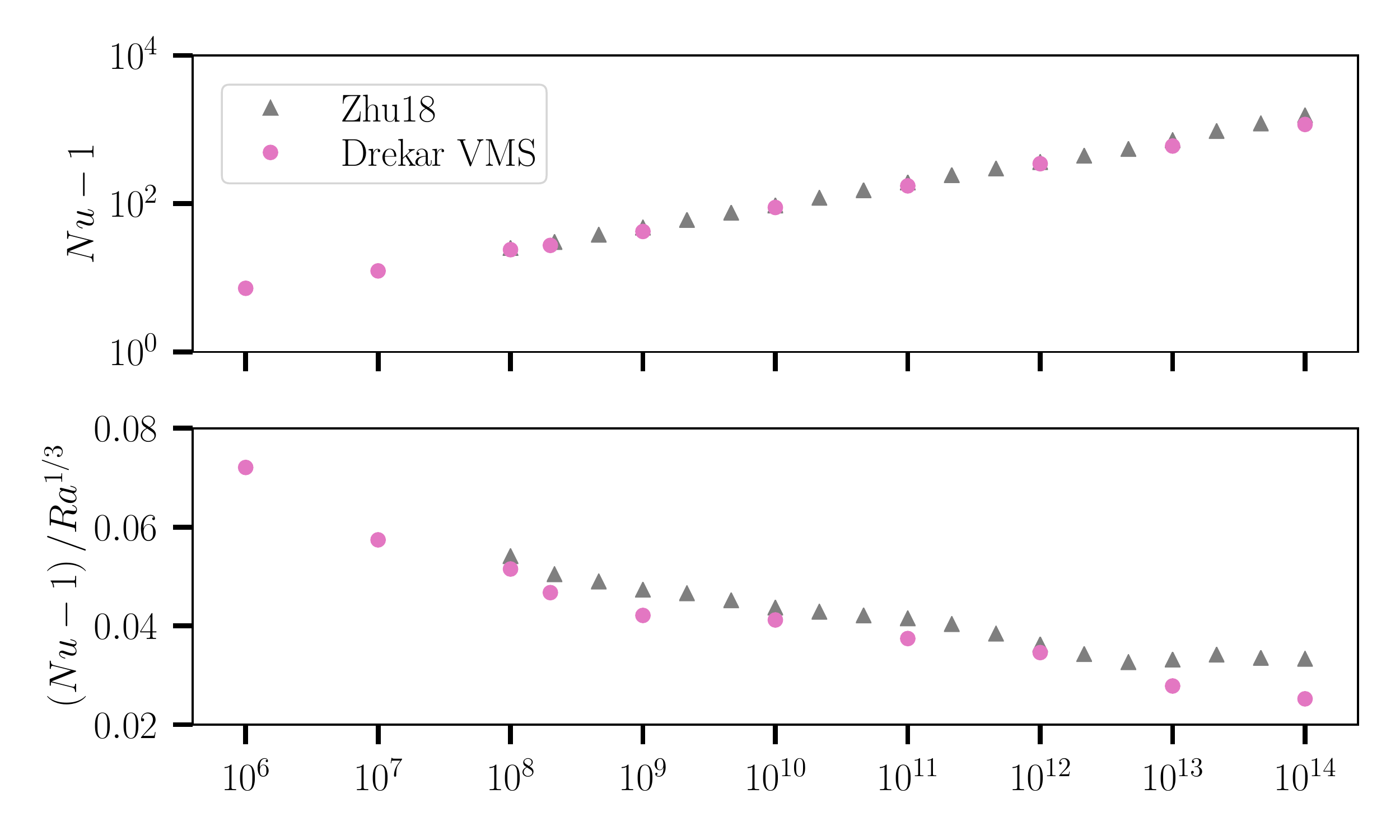}
  \caption{Top:  Nusselt-Rayleigh scaling of the 2D simulations showing good agreement between the VMS and DNS results.
Bottom:  Nusselt-Rayleigh scaling plot compensated by the classical $Ra^{1/3}$ scaling.  The VMS simulations show a
scaling closer to $2/7$ than $1/3$ while the DNS results show a transition to the $1/3$ scaling.}
  \label{fig:Nu_Ra_2D}
\end{figure}

The $Nu-Ra$ scaling for the 2D VMS simulations is compared to recent 2D DNS results in Figure~\ref{fig:Nu_Ra_2D}. 
The 2D VMS and DNS results are in excellent agreement up to $Ra\approx 10^{13}$ at which point the DNS results indicate an
increase in transport.  Recent work~\cite{zhu2018transition,zhu2019absence} has reported observations of a possible
transition to the ultimate regime in direct numerical simulations of two-dimensional \RB convection.  Although the present
VMS results do not indicate such a transition, we emphasize that the VMS simulations may be under-resolved in the
boundary layer.  Additional research is needed to determine the effectiveness of the VMS simulations for an under-resolved
boundary layer.  In this particular study, the 2D VMS simulations show $Nu = 0.151 Ra^{0.278}$ for $10^{10}\leq Ra \leq
10^{14}$.

\subsection{VMS-WALE Model Simulations of Three-dimensional \RB Convection}
\begin{figure}[h!]
  \centering
  \includegraphics[width=0.45\textwidth]{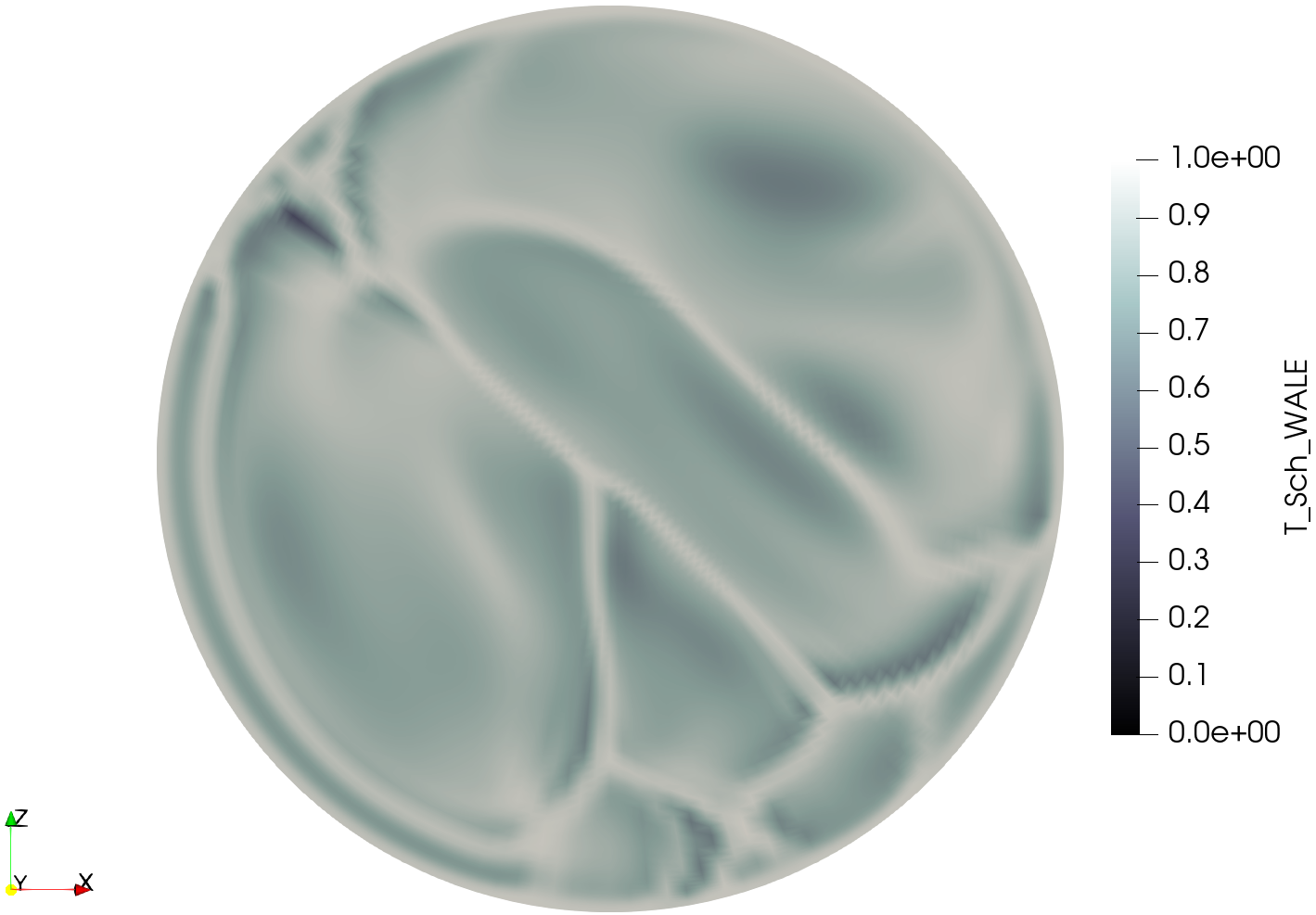}
  \caption{Temperature snapshot in a plane within the boundary layer at $y=0.001$ using a Schlieren visualization from the
VMS-WALE simulation in the cylinder geometry.}
  \label{fig:Ra100B_cylinder_plane_snapshot_WALE}
\end{figure}

\begin{figure}[h!]
  \centering
  \includegraphics[width=0.45\textwidth]{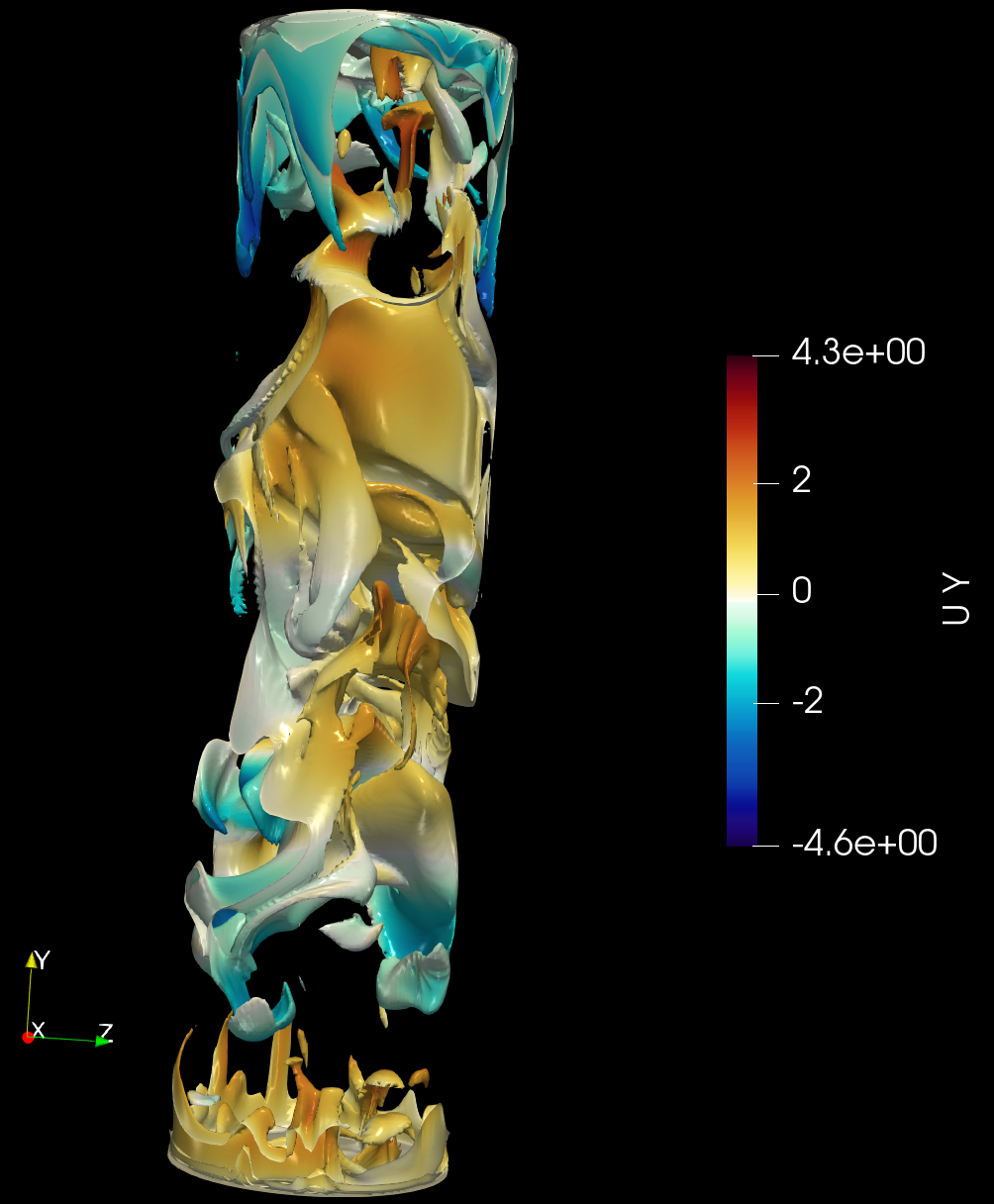}
  \caption{Temperature contours colored by vertical velocity at $Ra=10^{10}$.}
  \label{fig:Ra100B_cylinder_snapshot_WALE}
\end{figure}

The three-dimensional simulations were performed at $Pr=7$ in a circular cylinder of aspect ratio $1/4$.  The height of the
cylinder was $H=100$ for each case.  Homogeneous Neumann boundary conditions were used for the temperature on the surface of
the cylinder.  No-slip velocity boundary conditions were used on all surfaces of the cylinder.  Most of the 3D simulations
used the classical WALE model~\cite{nicoud1999subgrid, TMSMITH2011} as an eddy viscosity model in the VMS formulation with a
turbulent Prandtl number equal to unity.  This corresponds to the WALE-VMS model in Table~\ref{tab:mixed_params}.
Table~\ref{tab:3D_sim_params} provides a summary of the simulation parameters and results for the three-dimensional
simulations with the WALE-VMS model.
\begin{table}
  \begin{center}
\def~{\hphantom{0}}
  \begin{tabular}{cccccc}
      $Ra$           & $Nu$   & $N_{y}$ & $N_{r\theta}$ & Mesh Elements & $n_{f}$ \\[3pt]
       $10^{10}$     & 131.8  & 336     & 7344          & 2467584       & 106.4   \\
       $10^{11}$     & 267.6  & 256     & 4620          & 1182720       & 99.2    \\
       $10^{12}$     & 565.0  & 336     & 7344          & 2467584       & 442.8   \\
       $10^{13}$     & 1121.5 & 380     & 13500         & 5130000       & 90.4    \\
       $10^{14}$     & 2275.5 & 500     & 20800         & 10400000      & 37.78   \\
  \end{tabular}
  \caption{Simulation parameters and results for the three-dimensional runs.  $N_{y}$ and $N_{r\theta}$ represent the number
of elements in the vertical direction and in the $r-\theta$ plane, respectively.  $n_{f}$ represents the number of
free-fall times over which statistics were computed.}
  \label{tab:3D_sim_params}
  \end{center}
\end{table}
The results from the WALE-VMS model were compared to SUPG and VMS results using a sequence of finer meshes at $Ra=10^{10}$.
Table~\ref{tab:WALE-SUPG_VMS} compares the Nusselt number between these three models at the finest mesh.  At this mesh
resolution, all three models are in good agreement.
\begin{table}
  \begin{center}
\def~{\hphantom{0}}
  \begin{tabular}{cccccc}
      $Ra$           & Mesh Elements & VMS   & VMS-WALE & SUPG  \\[3pt]
      $10^{10}$      & 2,467,584     & 131.1 & 131.8    & 131.7 \\
  \end{tabular}
  \caption{Comparison of $Nu$ at $Ra=10^{10}$ between simulations in an aspect ratio $1/4$ cylinder using the VMS, VMS-WALE,
and SUPG models.}
  \label{tab:WALE-SUPG_VMS}
  \end{center}
\end{table}
Figure~\ref{fig:Ra100B_cylinder_plane_snapshot_WALE} shows a snapshot of the temperature field at $Ra=10^{10}$ taken at a plane in
the boundary layer ($y=0.001$) for the simulations in the cylinder.  Similarly to the 2D runs, the $y+$ value was computed
and determined to be less than $1$ for most runs.
Figure~\ref{fig:Ra100B_cylinder_snapshot_WALE} presents temperature contours colored by vertical velocity at $Ra=10^{10}$.

\begin{figure}[h!]
  \centering
  \includegraphics[width=0.45\textwidth]{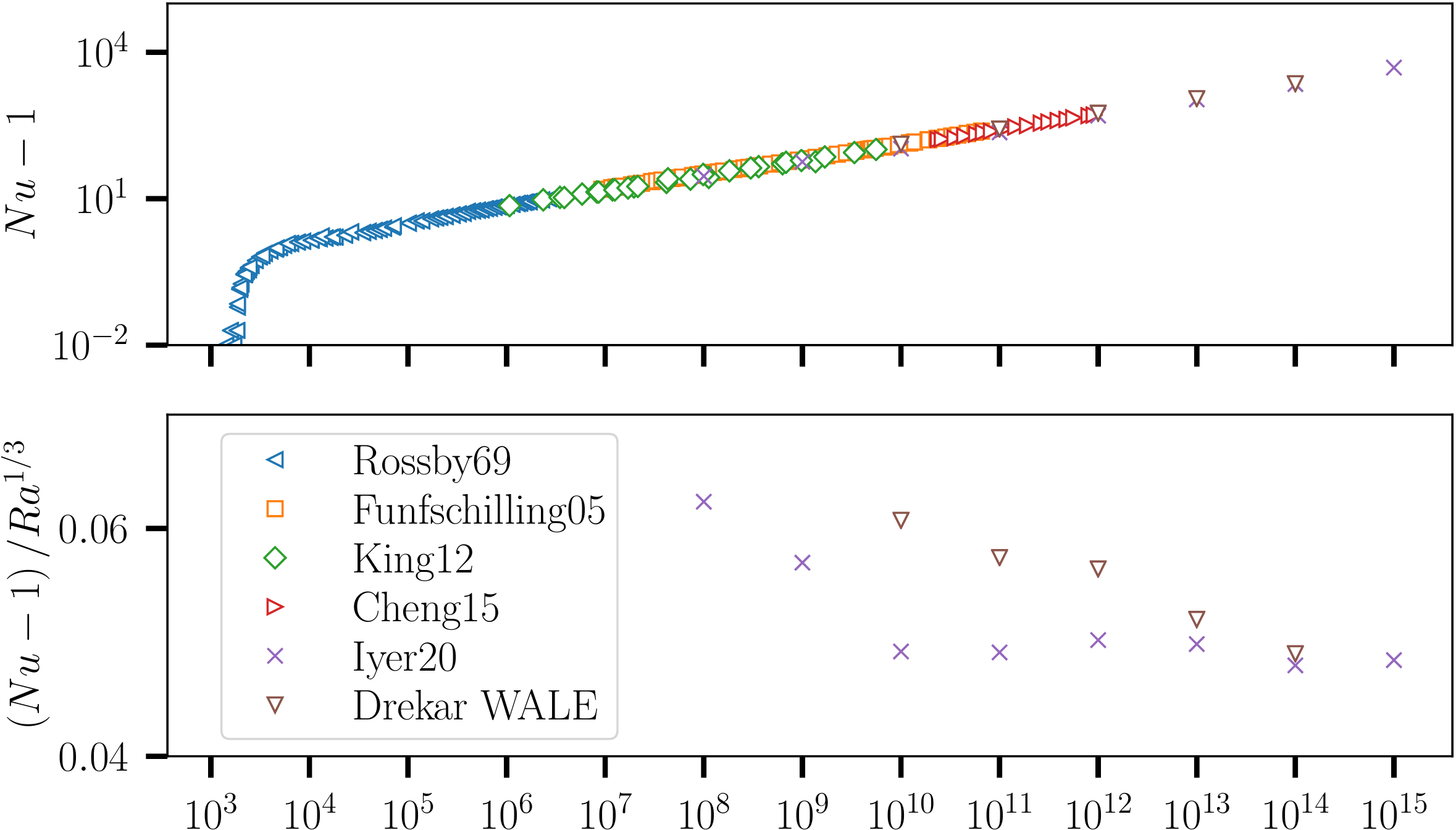}
  \caption{Nusselt-Rayleigh scaling for a variety of experimental and numerical studies: \cite{rossby1969study}---blue,
right-pointing triangles; \cite{funfschilling2005heat,nikolaenko2005heat}---orange squares; \cite{king2012heat}---green
diamonds; \cite{cheng2015laboratory}---red, right pointing triangles; \cite{iyer2020classical}---purple exes; this
work---brown, downward pointing triangles.}
  \label{fig:Nu_Ra}
\end{figure}

The Nusselt-Rayleigh scaling is presented in Figure~\ref{fig:Nu_Ra}.  The results are presented over several decades of $Ra$
for a number of studies, including the present results.
The results from the VMS models compare favorably to direct numerical simulations as well as experiments and the classical
$Nu\propto Ra^{1/3}$ scaling emerges over several decades in $Ra$.  In particular, the VMS-WALE model shows a scaling of $Nu
= 0.104 Ra^{0.310}$ for $10^{10}\leq Ra \leq 10^{14}$, while very recent DNS results show $Nu = 0.0525 Ra^{0.331}$ for
$10^{10}\leq Ra\leq 10^{15}$~\cite{iyer2020classical}.

\section{Conclusions}
\label{sec:conclusions}
A residual-based VMS formulation was derived for \RB convection and augmented with eddy viscosity models to account for the
Reynolds stresses.  This new mixed model was implemented in the finite element code Drekar.  In the current work, the WALE
model was used for the eddy viscosity model.  A number of two and three dimensional simulations were performed to compute the 
heat transport scaling in each system up to $Ra=10^{14}$.  The new VMS simulations are in good agreement with previous direct 
numerical simulations of two- and three- dimensional \RB convection results.  When compared to recent 2D DNS simulations, the 
VMS simulations begin to show a difference in $Nu-Ra$ scaling at around $Ra=10^{13}$.  Before claiming this as evidence for
the absence of a transition to the ultimate regime, we suggest several avenues for additional research regarding the VMS models.  

Resolution of the boundary layer in \RB convection is critical~\cite{grotzbach1983spatial}.  Residual-based VMS formulations 
have an ability to automatically adapt to regions of the flow that are under-resolved, but this may not be sufficient to
capture the underlying heat release from the boundary layer.  Moreover, when using linear finite elements, the viscous terms
in the residual vanish identically.  The impact of this incomplete residual on the $Nu-Ra$ scaling should be assessed.
Previous work has introduced techniques for reconstructing the diffusive flux for linear finite
elements~\cite{jansen1999better}.  We have implemented this diffusive flux reconstruction into the Drekar code and are
currently testing its impact on \RB convection.  In addition to the diffusive flux reconstruction, a more thorough mesh
convergence study should be performed along with a rigorous assessment of the near-wall behavior of the models.  Comparisons
to simulations that use higher-order elements would also provide a useful perspective.

Beyond \RB convection, we will implement and use VMS models on rotating \RB convection.  Early results from Drekar compare 
favorably with experiments in this regime.  Additional future work will include magnetoconvection with applications to
 geophysical and astrophysical problems.

\section*{Acknowledgments}
The authors would like to thank Professor Jon A. Aurnou (UCLA) for sharing data from their experimental system.
The work of John N. Shadid, Roger P. Pawlowski, Thomas Smith, and Sidafa Conde was partially supported by the U.S. Department
of Energy, Office of Science, Office of Advanced Scientific Computing Research. Sandia National Laboratories is a multi-mission
laboratory managed and operated by National Technology and Engineering Solutions of Sandia, LLC, a wholly owned subsidiary of 
Honeywell International, Inc., for the U.S. Department of Energy's National Nuclear Security Administration under contract
DE-NA0003525.  This paper describes objective technical results and analysis. Any subjective views or opinions that might be
expressed in the paper do not necessarily represent the views of the U.S. Department of Energy or the United States
Government.

\bibliographystyle{elsarticle-num}
\bibliography{refs,discretization_numerics}

\begin{thebibliography}{10}
\expandafter\ifx\csname url\endcsname\relax
  \def\url#1{\texttt{#1}}\fi
\expandafter\ifx\csname urlprefix\endcsname\relax\def\urlprefix{URL }\fi
\expandafter\ifx\csname href\endcsname\relax
  \def\href#1#2{#2} \def\path#1{#1}\fi

\bibitem{drazin2004hydrodynamic}
P.~G. Drazin, W.~H. Reid, Hydrodynamic stability, Cambridge university press,
  2004.

\bibitem{chandrasekhar2013hydrodynamic}
S.~Chandrasekhar, Hydrodynamic and hydromagnetic stability, Courier
  Corporation, 2013.

\bibitem{rayleigh1916lix}
L.~Rayleigh, {LIX}. on convection currents in a horizontal layer of fluid, when
  the higher temperature is on the under side, The London, Edinburgh, and
  Dublin Philosophical Magazine and Journal of Science 32~(192) (1916)
  529--546.

\bibitem{bodenschatz2000recent}
E.~Bodenschatz, W.~Pesch, G.~Ahlers, Recent developments in
  {R}ayleigh-{B}{\'e}nard convection, Annual review of fluid mechanics 32~(1)
  (2000) 709--778.

\bibitem{ahlers2009heat}
G.~Ahlers, S.~Grossmann, D.~Lohse, Heat transfer and large scale dynamics in
  turbulent {R}ayleigh-{B}{\'e}nard convection, Reviews of modern physics
  81~(2) (2009) 503.

\bibitem{malkus1954heat}
W.~V. Malkus, The heat transport and spectrum of thermal turbulence,
  Proceedings of the Royal Society of London. Series A. Mathematical and
  Physical Sciences 225~(1161) (1954) 196--212.

\bibitem{howard1966convection}
L.~N. Howard, Convection at high {R}ayleigh number, in: Applied Mechanics,
  Springer, 1966, pp. 1109--1115.

\bibitem{doering1996variational}
C.~R. Doering, P.~Constantin, Variational bounds on energy dissipation in
  incompressible flows. {III}. convection, Physical Review E 53~(6) (1996)
  5957.

\bibitem{ding2019exhausting}
Z.~Ding, R.~R. Kerswell, Exhausting the background approach for bounding the
  heat transport in {R}ayleigh-{B}\'{e}nard convection, arXiv preprint
  arXiv:1906.03376.

\bibitem{kraichnan1962turbulent}
R.~H. Kraichnan, Turbulent thermal convection at arbitrary {P}randtl number,
  The Physics of Fluids 5~(11) (1962) 1374--1389.

\bibitem{grossmann2000scaling}
S.~Grossmann, D.~Lohse, Scaling in thermal convection: a unifying theory,
  Journal of Fluid Mechanics 407 (2000) 27--56.

\bibitem{he2012transition}
X.~He, D.~Funfschilling, H.~Nobach, E.~Bodenschatz, G.~Ahlers, Transition to
  the ultimate state of turbulent {R}ayleigh-{B}{\'e}nard convection, Physical
  review letters 108~(2) (2012) 024502.

\bibitem{bouillaut2019transition}
V.~Bouillaut, S.~Lepot, S.~Auma{\^\i}tre, B.~Gallet, Transition to the ultimate
  regime in a radiatively driven convection experiment, Journal of Fluid
  Mechanics 861.

\bibitem{niemela2000turbulent}
J.~Niemela, L.~Skrbek, K.~Sreenivasan, R.~Donnelly, Turbulent convection at
  very high {R}ayleigh numbers, Nature 404~(6780) (2000) 837--840.

\bibitem{doering2019thermal}
C.~R. Doering, Thermal forcing and `classical' and `ultimate' regimes of
  {R}ayleigh-{B}{\'e}nard convection, Journal of Fluid Mechanics 868 (2019)
  1--4.

\bibitem{stevens2011prandtl}
R.~J. Stevens, D.~Lohse, R.~Verzicco, Prandtl and {R}ayleigh number dependence
  of heat transport in high {R}ayleigh number thermal convection, Journal of
  {F}luid {M}echanics 688 (2011) 31--43.

\bibitem{iyer2020classical}
K.~P. Iyer, J.~D. Scheel, J.~Schumacher, K.~R. Sreenivasan, Classical 1/3
  scaling of convection holds up to ${R}a= 10^{15}$, Proceedings of the
  National Academy of Sciences.

\bibitem{zhu2018transition}
X.~Zhu, V.~Mathai, R.~J. Stevens, R.~Verzicco, D.~Lohse, Transition to the
  ultimate regime in two-dimensional {R}ayleigh-{B}{\'e}nard convection,
  Physical {R}eview {L}etters 120~(14) (2018) 144502.

\bibitem{zhu2019absence}
X.~Zhu, V.~Mathai, R.~J. Stevens, R.~Verzicco, D.~Lohse, Absence of evidence
  for the ultimate regime in two-dimensional {R}ayleigh-{B}\'{e}nard convection
  reply, Physical review letters 123 (2019) 259402.

\bibitem{brooks1982streamline}
A.~N. Brooks, T.~J. Hughes, Streamline upwind/petrov-galerkin formulations for
  convection dominated flows with particular emphasis on the incompressible
  navier-stokes equations, Computer {M}ethods in {A}pplied {M}echanics and
  {E}ngineering 32~(1-3) (1982) 199--259.

\bibitem{franca2004stabilized}
L.~P. Franca, G.~Hauke, A.~Masud, Stabilized finite element methods,
  International Center for Numerical Methods in Engineering (CIMNE), Barcelona,
  Spain, 2004.

\bibitem{hughes1995multiscale}
T.~J. Hughes, Multiscale phenomena: Green's functions, the dirichlet-to-neumann
  formulation, subgrid scale models, bubbles and the origins of stabilized
  methods, Computer methods in applied mechanics and engineering 127~(1-4)
  (1995) 387--401.

\bibitem{masud2006multiscale}
A.~Masud, R.~Khurram, A multiscale finite element method for the incompressible
  {N}avier-{S}tokes equations, Computer Methods in Applied Mechanics and
  Engineering 195~(13-16) (2006) 1750--1777.

\bibitem{bazilevs2007variational}
Y.~Bazilevs, V.~Calo, J.~Cottrell, T.~Hughes, A.~Reali, G.~Scovazzi,
  Variational multiscale residual-based turbulence modeling for large eddy
  simulation of incompressible flows, Computer {M}ethods in {A}pplied
  {M}echanics and {E}ngineering 197~(1-4) (2007) 173--201.

\bibitem{liu2012residual}
J.~Liu, A.~Oberai, The residual-based variational multiscale formulation for
  the large eddy simulation of compressible flows, Computer {M}ethods in
  {A}pplied {M}echanics and {E}ngineering 245 (2012) 176--193.

\bibitem{sondak2015new}
D.~Sondak, J.~N. Shadid, A.~A. Oberai, R.~P. Pawlowski, E.~C. Cyr, T.~M. Smith,
  A new class of finite element variational multiscale turbulence models for
  incompressible magnetohydrodynamics, Journal of Computational Physics 295
  (2015) 596--616.

\bibitem{codina2000stabilized}
R.~Codina, On stabilized finite element methods for linear systems of
  convection--diffusion-reaction equations, Computer Methods in Applied
  Mechanics and Engineering 188~(1-3) (2000) 61--82.

\bibitem{codina2018variational}
R.~Codina, S.~Badia, J.~Baiges, J.~Principe, Variational multiscale methods in
  computational fluid dynamics, Encyclopedia of Computational Mechanics Second
  Edition (2018) 1--28.

\bibitem{xu2019residual}
S.~Xu, B.~Gao, M.-C. Hsu, B.~Ganapathysubramanian, A residual-based variational
  multiscale method with weak imposition of boundary conditions for
  buoyancy-driven flows, Computer Methods in Applied Mechanics and Engineering
  352 (2019) 345--368.

\bibitem{codina2002stabilized}
R.~Codina, Stabilized finite element approximation of transient incompressible
  flows using orthogonal subscales, Computer methods in applied mechanics and
  engineering 191~(39-40) (2002) 4295--4321.

\bibitem{wang2010spectral}
Z.~Wang, A.~Oberai, Spectral analysis of the dissipation of the residual-based
  variational multiscale method, Computer Methods in Applied Mechanics and
  Engineering 199~(13-16) (2010) 810--818.

\bibitem{hughes2005isogeometric}
T.~J. Hughes, J.~A. Cottrell, Y.~Bazilevs, Isogeometric analysis: {CAD}, finite
  elements, {NURBS}, exact geometry and mesh refinement, Computer methods in
  applied mechanics and engineering 194~(39-41) (2005) 4135--4195.

\bibitem{wang2010mixed}
Z.~Wang, A.~A. Oberai, A mixed large eddy simulation model based on the
  residual-based variational multiscale formulation, Physics of Fluids 22~(7)
  (2010) 075107.

\bibitem{oberai2014residual}
A.~A. Oberai, J.~Liu, D.~Sondak, T.~Hughes, A residual based eddy viscosity
  model for the large eddy simulation of turbulent flows, Computer Methods in
  Applied Mechanics and Engineering 282 (2014) 54--70.

\bibitem{nicoud1999subgrid}
F.~Nicoud, F.~Ducros, Subgrid-scale stress modelling based on the square of the
  velocity gradient tensor, Flow, {T}urbulence and {C}ombustion 62~(3) (1999)
  183--200.

\bibitem{jansen1999better}
K.~E. Jansen, S.~S. Collis, C.~Whiting, F.~Shaki, A better consistency for
  low-order stabilized finite element methods, Computer methods in applied
  mechanics and engineering 174~(1-2) (1999) 153--170.

\bibitem{shadid-jcp-10_mhd}
J.~N. Shadid, R.~P. Pawlowski, J.~W. Banks, L.~Chac\'on, P.~T. Lin, R.~S.
  Tuminaro, {Towards a scalable fully-implicit fully-coupled resistive {MHD}
  formulation with stabilized FE methods}, {J. Comput. Phys.} {229}~({20})
  ({2010}) {7649--7671}.

\bibitem{ShadidetalResistiveMHD2016}
J.~N. Shadid, R.~P. Pawlowski, E.~C. Cyr, R.~S. Tuminaro, L.~Chacon, P.~D.
  Weber, {Scalable Implicit Incompressible Resistive {MHD} with Stabilized FE
  and Fully-coupled {N}ewton-{K}rylov-{AMG}}, Comput. Methods Appl. Mech.
  Engrg. 304 (2016) 1--25.

\bibitem{OBER02}
C.~C. Ober, J.~N. Shadid, Studies on the accuracy of time-integration schemes
  for the radiation-diffusion equations, J. Comp. Phys. 195~(2) (2004)
  743--772.

\bibitem{ROPP03}
D.~L. Ropp, J.~N. Shadid, C.~C. Ober, Studies of the accuracy of time
  integration methods for reaction-diffusion equations, J. Comp. Phys. 194~(2)
  (2004) 544--574.

\bibitem{Ropp05}
D.~Ropp, J.~Shadid, Stability of operator splitting methods for systems with
  indefinite operators: reaction-diffusion systems, J. Comp. Phys. 203 (2005)
  449--466.

\bibitem{KnollWave05}
D.~A. Knoll, V.~A. Mousseau, L.~Chac\'on, J.~Reisner, Jacobian-free
  {N}ewton-{K}rylov methods for the accurate time integration of stiff wave
  systems, SIAM J. Sci. Comput. 25~(1) (2005) 213--230.

\bibitem{Ropp08}
D.~L. Ropp, J.~N. Shadid, Stability of operator splitting methods for systems
  with indefinite operators: {A}dvection-diffusion-reaction systems, J. Comp.
  Phys. 228 (2009) 3508--3516.

\bibitem{AscherPetzold98}
U.~M. Ascher, L.~R. Petzold, Computer Methods for Ordinary Differential
  Equations and Differential-Algebraic Equations, SIAM, 1998.

\bibitem{phipps2012ad}
E.~Phipps, R.~Pawlowski, Efficient expression templates for operator
  overloading-based automatic differentiation, in: S.~Forth, P.~Hovland,
  E.~Phipps, J.~Utke, A.~Walther (Eds.), Recent Advances in Algorithmic
  Differentiation, Vol.~87 of Lecture Notes in Computational Science and
  Engineering, Springer, Berlin, Heidelberg, 2012.

\bibitem{BS90}
P.~N. Brown, Y.~Saad, Hybrid {K}rylov methods for nonlinear systems of
  equations, SIAM J. Sci. Stat. Comput. 11 (1990) 450--481.

\bibitem{Knoll-Keyes}
D.~A. Knoll, D.~E. Keyes, Jacobian-free {N}ewton--{K}rylov methods: a survey of
  approaches and applications, J. Comput. Phys. 193 (2004) 357--397.

\bibitem{Lin06}
P.~T. Lin, M.~Sala, J.~N. Shadid, R.~S. Tuminaro, Performance of fully-coupled
  algebraic multilevel domain decomposition preconditioners for incompressible
  flow and transport, Int. J. Num. Meth. Eng. 67~(9) (2006) 208--225.

\bibitem{smith-ddbook}
B.~Smith, P.~Bjorstad, W.~Gropp, Domain Decomposition: Parallel Multilevel
  Methods for Elliptic Partial Differential Equations, Cambridge University
  Press, 1996.

\bibitem{Oosterlee}
U.~Trottenberg, C.~Oosterlee, A.~Sch\"uller, Multigrid, Academic Press, London,
  2001.

\bibitem{lin2009jcp}
P.~T. Lin, J.~N. Shadid, M.~Sala, R.~S. Tuminaro, G.~L. Hennigan, R.~J.
  Hoekstra, Performance of a parallel algebraic multilevel preconditioner for
  stabilized finite element semiconductor device modeling, Journal
  Computational Physics 228 (2009) 6250--6267.

\bibitem{LinShadidetalJCAM2017}
P.~T. Lin, J.~N. Shadid, J.~J. Hu, R.~P. Pawlowski, E.~C. Cyr, Performance of
  fully-coupled algebaric multigrid preconditioners for large-scale {VMS}
  resistive {MHD}, J. Comp. and Applied Math 344 (2018) 782--793.

\bibitem{Dennis-Schnabel}
J.~E. Dennis, Jr., R.~B. Schnabel, Numerical Methods for Unconstrained
  Optimization and Nonlinear Equations, Series in Automatic Computation,
  Prentice-Hall, Englewood Cliffs, NJ, 1983.

\bibitem{EW96}
S.~C. Eisenstat, H.~F. Walker, Choosing the forcing terms in an inexact
  {N}ewton method, SIAM J. Sci. Comput. 17 (1996) 16--32.

\bibitem{JN99}
J.~N. Shadid, A fully-coupled {N}ewton-{K}rylov solution method for parallel
  unstructured finite element fluid flow, heat and mass transfer simulations,
  Int. J. Comput. Fluid Dynamics 3-4 (1999) 199--211.

\bibitem{ml-guide}
M.~Gee, C.~Siefert, J.~Hu, R.~Tuminaro, M.~Sala, {ML} 5.0 smoothed aggregation
  user's guide, Tech. Rep. SAND2006-2649, Sandia National Laboratories,
  Albuquerque NM, 87185 (2006).

\bibitem{tuminaro2002}
R.~Tuminaro, C.~Tong, J.~Shadid, K.D.Devine, D.~Day, On a multilevel
  preconditioning module for unstructured mesh {K}rylov solvers: two-level
  schwarz, Comm. Num. Method. Eng. 18 (2002) 383--389.

\bibitem{shadid2004}
J.~Shadid, R.~Tuminaro, K.~Devine, G.~Henningan, P.~Lin, Performance of
  fully-coupled domain decomposition preconditioners for finite element
  transport/reaction simulations, J. Comput. Phys. 205~(1) (2005) 24--47.

\bibitem{shadid2006}
J.~N. Shadid, A.~G. Salinger, R.~P. Pawlowski, P.~T. Lin, G.~L. Hennigan, R.~S.
  Tuminaro, R.~B. Lehoucq, Stabilized {FE} computational analysis of nonlinear
  steady state transport/reaction systems, Comp. Meth. Applied Mech. Eng. 195
  (2006) 1846--1871.

\bibitem{TMSMITH2011}
T.~M. Smith, J.~N. Shadid, R.~P. Pawlowski, E.~C. Cyr, P.~D. Weber, Reactor
  core sub-assembly simulations using a stabilized finite element method, in:
  NURETH-14-500, The 14th International Topical Meeting on Nuclear Reactor
  Thermalhydraulics, NURETH-14, Toronto, Ontario, Canada, 2011.

\bibitem{rossby1969study}
H.~Rossby, A study of {B}{\'e}nard convection with and without rotation,
  Journal of Fluid Mechanics 36~(2) (1969) 309--335.

\bibitem{funfschilling2005heat}
D.~Funfschilling, E.~Brown, A.~Nikolaenko, G.~Ahlers, Heat transport by
  turbulent {R}ayleigh-{B}{\'e}nard convection in cylindrical samples with
  aspect ratio one and larger, Journal of Fluid Mechanics 536 (2005) 145--154.

\bibitem{nikolaenko2005heat}
A.~Nikolaenko, E.~Brown, D.~Funfschilling, G.~Ahlers, Heat transport by
  turbulent {R}ayleigh-{B}{\'e}nard convection in cylindrical cells with aspect
  ratio one and less, Journal of Fluid Mechanics 523 (2005) 251--260.

\bibitem{king2012heat}
E.~M. King, S.~Stellmach, J.~M. Aurnou, Heat transfer by rapidly rotating
  {R}ayleigh-{B}{\'e}nard convection, Journal of Fluid Mechanics 691 (2012)
  568--582.

\bibitem{cheng2015laboratory}
J.~S. Cheng, S.~Stellmach, A.~Ribeiro, A.~Grannan, E.~M. King, J.~M. Aurnou,
  Laboratory-numerical models of rapidly rotating convection in planetary
  cores, Geophysical Journal International 201~(1) (2015) 1--17.

\bibitem{grotzbach1983spatial}
G.~Gr{\"o}tzbach, Spatial resolution requirements for direct numerical
  simulation of the {R}ayleigh-{B}{\'e}nard convection, Journal of
  computational physics 49~(2) (1983) 241--264.

\end{thebibliography}

\end{document}